\documentclass[aps,twocolumn,showpacs,preprintnumbers,nofootinbib,prd,10pt,superscriptaddress]{revtex4-1}

\makeatletter
\def\l@subsubsection#1#2{}
\def\l@subsubsubsection#1#2{}
\makeatother

\usepackage{graphicx,amssymb,amsmath,amsthm,amsfonts,epsfig,epsf,fixmath}
\usepackage{comment}
\usepackage{mathrsfs}
\usepackage[usenames]{color}
\usepackage{epstopdf}
\usepackage{aas_macros}
\usepackage{bm}
\usepackage{dcolumn}
\usepackage{latexsym}
\usepackage{rotating}
\usepackage{longtable}

\usepackage{enumerate}
\usepackage{tensor,multirow}
\usepackage{url}
\usepackage[dvipsnames]{xcolor}
\usepackage[unicode]{hyperref}
\hypersetup{colorlinks=true, citecolor=MidnightBlue,
            linkcolor=MidnightBlue, urlcolor=MidnightBlue, linktocpage=true}

\newcommand{\tn}{\textnormal}

\newcommand{\dd}{\mathrm{d}}

\newcommand{\source}{\mathcal{T}^\omega_{\ell m}}

\newcommand{\be}{\begin{equation}}
\newcommand{\eeq}{\end{equation}}
\newcommand{\ba}{\begin{align}}
\newcommand{\ea}{\end{align}}
\newcommand{\dE}{\dot E}
\newcommand{\dL}{\dot L}

\newcommand{\dC}{\dot{\mathcal{C}}}

\newcommand{\hcross}{h^{\times}}

\newcommand{\braket}[2]{\langle #1 | #2 \rangle}
\renewcommand{\l}{\ell}
\newcommand{\GSSI}{Gran Sasso Science Institute (GSSI), I-67100 L’Aquila, Italy}
\newcommand{\GranSasso}{INFN, Laboratori Nazionali del Gran Sasso, I-67100 Assergi, Italy}

\begin{document}
\title{Extreme mass-ratio inspirals as probes of scalar fields: \\ inclined circular orbits around Kerr black holes}

\author{Matteo Della Rocca}
\email{matteo.dellarocca@phd.unipi.it}
\affiliation{Dipartimento di Fisica, Universit\`a di Pisa, Largo B. Pontecorvo 3, 56127 Pisa, Italy}
\affiliation{INFN, Sezione di Pisa, Largo B. Pontecorvo 3, 56127 Pisa, Italy}
\author{Susanna Barsanti}
 \email{susanna.barsanti@uniroma1.it}
\affiliation{Dipartimento di Fisica, ``Sapienza'' Universit\`a di Roma, Piazzale 
Aldo Moro 5, 00185, Roma, Italy}
\affiliation{Sezione INFN Roma1, Roma 00185, Italy}
\author{Leonardo Gualtieri}
 \email{leonardo.gualtieri@unipi.it}
\affiliation{Dipartimento di Fisica, Universit\`a di Pisa, Largo B. Pontecorvo 3, 56127 Pisa, Italy}
\affiliation{INFN, Sezione di Pisa, Largo B. Pontecorvo 3, 56127 Pisa, Italy}
\author{Andrea Maselli}
 \email{andrea.maselli@gssi.it}
\affiliation{\GSSI}
\affiliation{\GranSasso}

\begin{abstract} 
Extreme mass-ratio inspirals, a target source for the space-based gravitational wave detector LISA, are a sensitive probe of fundamental scalar fields coupled to gravity.
We assess the capability of LISA to detect whether the secondary compact object is endowed with a scalar field, in the case of inclined orbits. We show that the imprint of the scalar field depends on the orbital inclination, and is significantly larger for prograde orbits. 
\end{abstract}

\maketitle

\section{Introduction}
%

Asymmetric binaries with mass ratios $q\ll1$ represent a new family of gravitational wave (GW) sources, that merge in a frequency band dim to current interferometers. Assembled by a massive black hole (BH) (the \textit{primary}) and by a lighter stellar mass object (the \textit{secondary}), either a BH or a neutron star, such systems typically emit GWs at frequencies below 1 Hz\footnote{Exotic configurations with a sub-solar mass secondary inspiralling around an intermediate mass BH could also provide a new type of EMRI for 3G ground-based detectors \cite{Barsanti:2021ydd}.}. 
Among asymmetric binaries, Extreme Mass Ratio Inspirals (EMRIs) with a primary mass $M\gtrsim 10^5M_{\odot}$ and $q<10^{-4}$ feature unique dynamical properties, coalescing in the mHz regime, with a GW emission peaking into the bucket of the LISA sensitivity curve \cite{Berry:2019wgg}.

EMRI evolution is mostly dictated by their mass ratio, with the duration of the inspiral and the number of GW cycles growing as $q$ decreases \cite{Barack:2018yvs}, allowing such sources to stay in the LISA band for hundreds of thousands of orbits. 

The large number of GW cycles performed on a highly relativistic dynamics, supplied by the extreme variability of the orbital evolution, promise measurements of the source parameters with unparalleled accuracy \cite{Berry:2019wgg}. 
Such properties render EMRIs golden targets to probe a variety of fundamental physics science cases \cite{Cardenas-Avendano:2024mqp}. 
These include precise tests of General Relativity (GR)~\cite{Barack:2006pq,Babak:2017tow}, of the multipolar structure of compact objects~\cite{Barack:2006pq,Babak:2017tow,Fransen:2022jtw,Raposo:2018xkf,Bena:2020see,Bianchi:2020bxa,Loutrel:2022ant,Piovano:2020ooe,Pani:2019cyc,Piovano:2022ojl}, searches of new physics at the horizon-scale physics~\cite{Datta:2019epe,Datta:2019euh,Maggio:2021uge}, of the existence of exotic compact objects~\cite{Pani:2010em,Macedo:2013qea,Destounis:2023khj,Datta:2019epe,Datta:2019euh,Maggio:2021uge}, and of new fundamental fields coupled to the gravity sector \cite{Cardoso:2011xi,Yunes:2011aa,Pani:2011xj,Canizares:2012is,Hannuksela:2018izj,Hannuksela:2019vip,Maselli:2020zgv,Maselli:2021men,Barsanti:2022ana,Barsanti:2022vvl,Liang:2022gdk,Zhang:2023vok,Zi:2022hcc,Lestingi:2023ovn,Collodel:2021jwi}. 

EMRIs are suitably described through relativistic perturbation theory, exploiting the small mass ratio $q$ as natural parameter for the expansion. Taking advantage of this setup, the Self-Force (SF) approach provides the best method to model EMRIs \cite{Barack:2018yvs}. 
Parameter estimation requirements ask for waveform templates accurate at the post-adiabatic order, i.e. yielding a $\mathcal{O}(q)$ phase error over the course of the inspiral. Developing such models in GR has provided a formidable challenge. The first post-adiabatic waveforms have been developed only recently, for quasi-circular inspirals around Schwarzschild BHs \cite{Pound:2019lzj, Warburton:2021kwk, Wardell:2021fyy}. 
Current efforts aim to improve such models in order to describe BHs on inclined, eccentric orbits, taking into account spin effects for both the primary and the secondary \cite{Green:2019nam,Dolan:2021ijg,Upton:2021oxf,Toomani:2021jlo,Osburn:2022bby,Spiers:2023cip,Nasipak:2021qfu,Piovano:2020zin,Mathews:2021rod,Drummond:2022xej,Upton:2023tcv,Drummond:2023loz,Upton:2023tcv}.

Exploiting the full EMRIs potential to test gravity and detect new fundamental fields requires accurate waveforms to be compared against data. However, EMRI modelling beyond GR is at its infancy, with the complexity of calculations growing fast because of the new fields and their couplings. This picture is worsened by the lack of a Kerr-like solution to use as a background for the perturbations.
Only recently, Refs. ~\cite{Li:2022pcy,Wagle:2023fwl} proposed a new formalism to derive a generalization of Teukolsky's equation in modified theories of gravity.

So far, the vast majority of studies has focused on assessing the relevance of EMRI observations to probe the spacetime around the \textit{primary} BH \cite{Barack:2006pq,Glampedakis:2005cf}. However, it was recently shown that, for a wide class of gravity theories with non-minimally coupled scalar fields, the scalar charge of the \textit{secondary} could leave a significant imprint on the EMRI emission, measurable with  exquisite precision by LISA \cite{Maselli:2020zgv}. 
Working in an Effective Field Theory approach it was also pointed out that, in such theories, the scalar charge of the primary is negligible at the leading order in $q$. This leads to drastic simplifications for the EMRI treatment beyond GR, with the primary being adequately described by the Kerr metric, and the deviations from GR fully controlled by the charge of the secondary\footnote{This approach was also generalised to study the spectrum of quasi-normal modes for massive BHs in shift-symmetric scalar tensor theories \cite{DAddario:2023erc}.}. 
More recently, this framework was framed into a rigorous SF scheme, developing a consistent formalism to compute perturbations at the first and second-order in the mass ratio, and derive post-adiabatic waveform corrections \cite{Spiers:2023cva}. 

Within this approach some of us studied the adiabatic evolution of EMRIs with massless scalar fields on equatorial circular \cite{Maselli:2020zgv,maselli_detecting_2022} (hereafter paper I and II, respectively) and eccentric orbits \cite{Barsanti:2022ana} (paper III), investigating the relevance of the secondary charge on the binary dynamics, and its detectability by LISA. 
Motivated by the complex orbital configurations expected for EMRIs, in this paper we make a step forward and study the GW emission of such systems on inclined circular trajectories. We evolve binaries with different charges, and assess the detectability of the scalar charge by LISA observations, as a function  of the orbital inclination.

The rest of the paper is organized as follows. In Sec.~\ref{sec:theory} we describe the theoretical  setup for modelling EMRIs  with circular, inclined orbits, in the presence of massless scalars; we derive the main equations and discuss the numerical implementation needed to compute the GW fluxes. In Sec.~\ref{sec:results} we assess the relevance of orbital inclination  on the distinguishability 
between waveforms with an without the additional scalar charge. Conclusions and future prospects are discussed in Sec.~\ref{sec:conclusions}.

%
\section{EMRIs and scalar fields: the theoretical minimum}\label{sec:theory}
%
In this Section we briefly recall the theoretical background of our approach; for further details see Papers I-III. 
We use geometrical ($G=c=1$) units.
%
\subsection{Massless scalar fields in the Kerr spacetime}\label{subsec:fields}
%
We consider a general action of the form (see Papers I,II and\,\cite{Spiers:2023cva}):

\begin{equation}
    S\left[\textbf{g}, \varphi, \Psi \right] = S_0\left[\textbf{g}, \varphi\right] + \alpha S_{\rm c} \left[\textbf{g}, \varphi\right] + S_{\rm m}\left[\textbf{g}, \varphi, \Psi\right]\ ,
    \label{action}
\end{equation}
where ${\bf g}$  is the spacetime metric, $\varphi$ is a real, massless scalar field,
\begin{equation}    
S_0 = \int \dd^4 x \frac{\sqrt{-g}}{16 \pi} \left(R - \frac{1}{2} \partial_\mu \varphi \partial^{\mu} \varphi \right)\,
\end{equation}
and $R$ is the Ricci scalar. 
The coupling between the scalar field and the metric is encoded in the action $\alpha S_{\rm c}$, which we assume to be analytic 
in $\varphi$.

Our formalism can be easily extended to massive scalars, as discussed in~\cite{barsanti_detecting_2022}.
We assume that the coupling constant $\alpha$ has dimensions $({\rm mass})^n$, with $n>1$, namely that the interactions are suppressed by some characteristic energy scale (in physical units). Matter fields, denoted by $\Psi$, are described by the action $S_{\rm m}$.

The action\,\eqref{action} yields the field equations
\begin{equation}
    G_{\mu\nu} =   8 \pi T^{\rm scal}_{\mu \nu}+\alpha T^{\rm c}_{\mu\nu}+ T^{\rm m}_{\mu\nu}\quad\ ,\quad \Box \varphi  = T^{\rm c}+ T^{\rm m} \ ,\label{eq:fieldseq}
\end{equation}
where $\square=\nabla_\mu\nabla^\mu$,
$T^{\rm scal}_{\mu \nu} = \frac{1}{16 \pi } \left[\partial_{\mu} \varphi\partial_{\nu} \varphi - \frac{1}{2}g_{\mu\nu}(\partial \varphi)^2\right]$ 
and 
\begin{equation}
T^{\rm c}_{\mu\nu} =- \frac{16 \pi}{\sqrt{-g}}\frac{\delta S_{\rm c}}{\delta g^{\mu \nu}}   \quad\ ,\quad
T^{\rm m} =- \frac{16 \pi}{\sqrt{-g}}\frac{\delta S_{\rm m}}{\delta \varphi}\ . 
\end{equation}
We shall now discuss the key simplifications that occur for EMRIs, and allow to disentangle tensor and scalar perturbations  at the leading dissipative order. We refer the reader to \cite{Spiers:2023cva} for further details, as well as for the the extension of such formalism to post-adiabatic corrections.

We consider binaries in which the primary is a BH of mass $M$, with the latter being the only physical scale of the background. Hence, since we assume that for $\alpha\rightarrow 0$ solutions of Eqs.~\eqref{eq:fieldseq} are continuously connected to GR solutions, deviations from the latter must depend on\footnote{In our units both the metric and the scalar field are dimensionless.}
\begin{equation}
\zeta=\frac{\alpha}{M^n}=q^n\frac{\alpha}{m^n_{\rm p}}\ ,\label{eq:zetapar}
\end{equation}
where $m_p$ is the mass of the EMRI secondary. Astrophysical constrains imply $\alpha/m_p^n\sim \mathcal{O}(1)$ and smaller \cite{Nair:2019iur}, such that $\zeta\ll 1$. This allows us to exploit $q$, the natural parameter used to describe EMRIs within the perturbative self-force (SF) approach in GR, as a single bookkeeping parameter for our physical setup. As shown in \cite{Spiers:2023cva}, by expanding the fields equations, the metric and the scalar field in powers of $q$,
\begin{align}
g_{\mu\nu}=g_{\mu\nu}^{(0)}+q h^{(1)}_{\mu\nu}+\ldots\ \ ,  
\ \varphi=\varphi^{(0)}+q \varphi^{(1)}+\ldots\ ,   
\end{align} 
we can define a SF scheme for the EMRI evolution. In this paper we focus on the leading dissipative contribution, which is fully determined by $h_{\mu\nu}^{(1)}$ and $\varphi^{(1)}$.

At the zero order in the mass ratio, the background spacetime is described by the Kerr metric. The scalar field $\varphi^{(0)}$, whose contribution arises from $S_0$, is constant due to no-hair theorems \cite{1970CMaPh..19..276C, bekenstein_novel_1995, hawking_black_1972, Sotiriou:2011dz, Hui:2012qt}, and can be set to zero without loss of generality. 

At first order in $q$, metric and scalar field perturbations are sourced by the presence of the secondary, which we describe using the so-called skeletonized approach \cite{1975ApJ...196L..59E,Damour:1992we}, in which the matter action $S_{\rm m}$ is replaced by a point particle action $S_{\rm p}$. For a massive, scalar-charged, compact object:
\begin{equation}
    S_{\rm p} = - 
    \int_\gamma m \left(\varphi \right) \sqrt{g_{\mu \nu} \frac{\dd y^\mu_{\rm p}}{\dd\lambda}\frac{\dd y^\nu_{\rm p}}{\dd\lambda}} 
    \dd\lambda\ ,\label{eq:partac}
\end{equation}
where $\gamma$ is the worldline of the particle, with four velocity $dy_p/d\lambda$ and proper time $\lambda$. Eq.\,\eqref{eq:partac} depends on the mass function $m(\varphi)$, which sources the scalar charge of the secondary, $d$ \cite{Julie:2017ucp,Julie:2017rpw}. 
The latter is determined by expanding $\varphi$ in a buffer region inside the world-tube containing the stellar mass object, 
\begin{equation}
    \varphi^{(1)}=\frac{m_{\rm p} d}{{\tilde r}}
    +O\left(m^2_{\rm p}/{\tilde r}^2\right)\ ,
    \label{exp_sc}
\end{equation}
where $\{\tilde{x}_{\mu}\}$ is a reference frame centered on the secondary and the distance $\tilde r$ from the worldline is such that $m_p\ll \tilde{r}\ll M$. By replacing the solution\,\eqref{exp_sc} in the field equation for the scalar fields, one finds the matching conditions $m_{\rm p}=m(0)$, and $d=-4m'(0)/m_{\rm p}$. 

Expanding Eqs.~\eqref{eq:fieldseq} at the linear order in $q$, supplied by the action $S_p$, yields a set of decoupled equations for the metric and the scalar field perturbation:
\begin{align}
    G^{\alpha \beta}[h^{(1)}_{\alpha\beta}] &= 8 \pi m_{\rm p}  \int \frac{\delta^{(4)}\left(x-y_p(\lambda)\right)}{\sqrt{-g}}\frac{\dd y^{\alpha}_p}{\dd\lambda} \frac{\dd y^{\beta}_p}{\dd\lambda} \dd\lambda \ ,\label{einstein}\\
    \Box \varphi^{(1)} &= - 4 \pi d m_{\rm p} \int \frac{\delta^{(4)} \left(x - y_p(\lambda)\right) }{\sqrt{-g}}\dd\lambda\ . 
    \label{scal_schw}
\end{align}
The amplitude of $\varphi^{(1)}$ is controlled by the value of the scalar charge.

Equations \eqref{einstein}-\eqref{scal_schw} have been solved in Papers I-III 
for circular and eccentric equatorial orbits, in order to compute the emitted energy and angular momentum fluxes.

%
\subsection{Non-equatorial, circular geodesics of Kerr spacetime}\label{subsec:geodesics}
%
We focus on EMRIs moving on geodesics of the Kerr spacetime, the latter being described, in Boyer-Lindquist coordinates $x^\mu=(t,r,\theta,\phi)$, by the following line element:
\begin{equation}
\begin{split}
        \dd s^2=-\left(1-\frac{2Mr}{\Sigma}\right)\dd t^2-\frac{4Mra\sin^2\theta}{\Sigma}\dd t \dd\phi+\Sigma \dd\theta^2 +\\
    +\frac{\Sigma}{\Delta}dr^2+\left(r^2+a^2+\frac{2Mra^2}{\Sigma}\sin^2\theta\right)\sin^2\theta \dd\phi^2.
\label{KerrMetricBL}
\end{split}
\end{equation}
where $M$ and $a$ are the BH mass and spin parameter, while $\Delta= r^2+a^2-2 M r$ and $\Sigma= r^2+a^2 \cos^2\theta$.

The geodesic equations for $(t,r,\phi, \theta)$ are given by:
\begin{align}
\begin{split}
     \Sigma\frac{\dd t}{\dd\tau}&=E\left[\frac{(r^2+a^2)^2}{\Delta}-a^2\sin^2\theta\right]+\\
     &+aL\left(1-\frac{r^2+a^2}{\Delta}\right)\label{tGeoEquation},
\end{split}\\
    \Sigma\frac{\dd\phi}{\dd\tau}&=\frac{L^2}{\sin^2\theta}+aE\left(\frac{r^2+a^2}{\Delta}-1\right)-\frac{a^2L}{\Delta}\label{VarphiGeoEquation},\\
\begin{split}
       \left(\Sigma\frac{\dd r}{\dd \tau}\right)^2&=\mathcal{R}(r)=-\Delta\left[r^2+(L-aE)^2+Q\right]\\
       &+\left[E(r^2+a^2)-La\right]^2\,.\label{RGeoEquation_general_case}
\end{split}\\
    \left(\Sigma\frac{\dd\theta}{\dd\tau}\right)^2&=\Theta^2(\theta)= Q-\cos^2\theta\left[(1-E^2)a^2+\frac{L^2}{\sin^2\theta}\right]\label{ThetaGeoEquation}\,,
\end{align}
where $\tau$ is the proper time, $E$ and $L$ are the energy and angular momentum of the particle per unit mass at infinity, respectively, and $Q$ is the Carter constant. We focus on  bound orbits, for which $0\le E<1$ and $Q \ge 0$. 

In the orbital motion, the polar angle oscillates between $\theta_{\tn{min}}$ and $\theta_{\tn{max}}=\pi-\theta_{\tn{min}}$; the value of $\theta_{\tn{min}}$ is given by the equation $\Theta^2(\theta_{\tn{min,max}})=0$, which can be cast as an algebraic quadratic equation by changing variable to  $z=\cos^2(\theta)$. Its solutions are $z_-,z_+$, with  $z_- \le  z_+$; note that $\cos^2\theta_{\tn{min}}=\cos^2\theta_{\tn{max}}=z_-$, while $z_+>1$\,\cite{chandrasekhar_mathematical_2009}. 

It is useful to perform a further change of variable, by introducing the angular variable $\chi$, such that
\begin{equation}
    z=\cos^2\theta= z_-\cos^2\chi\,. 
\label{chi_Definition}
\end{equation}
A period of the variable $\theta$, from $\theta_{\tn{min}}$ to $\theta_{\tn{max}}$ and back, corresponds to a period  $[0,2\pi]$ of $\chi$; indeed, $\chi(\theta_{\tn{min}})=0,2\pi$, $\chi(\theta_{\tn{max}})=\pi$.

We shall consider a circular geodesic, at a $r=r_0$ constant. Note that, as shown in\,\cite{kennefick_radiation-reaction-induced_1996}, circular orbits in Kerr spacetime remain circular during the inspiral. Indeed, the time derivative of the eccentricity is vanishing for circular orbits evolving in the adiabatic regime, and can be neglected.
This proof applies for a generic external force in the Kerr background and easily extends to the case of an additional radiating scalar field. 

The geodesic equations\,\eqref{tGeoEquation}-\eqref{VarphiGeoEquation}, in terms of the variable $\chi$, reduce to:
\begin{align}
    \frac{\dd t}{\dd\chi}&=\frac{\gamma+a^2 E z(\chi)}{\sqrt{\beta(z_+-z(\chi))}}
\label{t_Derivative_chi}\\
\frac{\dd\phi}{\dd\chi}&=\frac{1}{\sqrt{\beta(z_+-z(\chi))}}\left[\frac{L}{1-z(\chi)}+\delta\right],
\label{Phi_Derivative_chi}
\end{align}
where
\begin{align}
\beta=& a^2(1-E^2),\\
\gamma=& E\left[\frac{(r_0^2+a^2)^2}{\Delta_0}-a^2\right]+aL\left(1-\frac{r_0^2+a^2}{\Delta_0}\right) \label{Gamma_Definition},\\
\delta=&aE\left(\frac{r_0^2+a^2}{\Delta_0}-1\right)-\frac{a^2L}{\Delta_0}\ ,
\end{align}
and $\Delta_0=r_0^2+a^2-2Mr_0$.  Eqs.\,\eqref{t_Derivative_chi}-\eqref{Phi_Derivative_chi} 
can be integrated using elliptic functions (see Appendix\,\ref{app:elliptic}).

For circular equatorial trajectories, the orbital motion is described by the natural fundamental frequency $\dd\phi/\dd t$. For inclined orbits, the picture is more complex since $\dd\phi/\dd t$ depends on $\theta$. In our setup we can define two fundamental frequencies, $\Omega_{\phi}$  and $\Omega_\theta$ as follows. We define the polar period $T_\theta=t(2\pi)=4t({\pi}/{2})$.
This is the time interval in which $\chi$ varies from $0$ to $2\pi$ (and $\theta$ from
$\theta_{\tn{min}}$ to $\theta_{\tn{max}}$ and back to $\theta_{\tn{min}}$). In terms of $T_\theta$, we define the polar frequency
$\Omega_\theta={2\pi}/{T_\theta}$, and  
$\bar{\phi}= \phi(T_{\theta})=4\phi\left({\pi}/{2}\right)$; note that for a rotating BH, $\bar\phi\neq2\pi$. Finally, we define the azimuthal frequency $\Omega_\phi=\bar\phi/T_\theta$.

As shown by Eq.~\eqref{Phi_Derivative_chi}, $f=\dd\phi/\dd t$ 
depends on the polar angle $\theta$ only, namely it is periodic 
in time with period $T_{\theta}$. Then, it can be decomposed 
as a Fourier series
\begin{equation}
    f(\theta)=\sum_{n=-\infty}^\infty   f_{n}e^{in\Omega_{\theta}t} \ , \ f_{n}=\frac{1} {T_\theta}\int_{0}^{T_{\theta}}dt \ f(\theta)e^{-in\Omega_{\theta}t}.
\label{Fourier_Series_Periodic_Function_Theta}
\end{equation}
By integration we obtain 
\begin{equation}
    \phi(t)=\Omega_\phi t+\sum_{n\in \mathbb{Z}\setminus\{0\}}a_{n}e^{in\Omega_\theta t}\ ,
\label{phi_thetaHarmonics_decomposition}
\end{equation}
where $a_n=-i f_n/(n\Omega_{\theta}), \ n\ne 0$.
\subsection{Adiabatic inspirals}
\label{sec:perturbations}
%
%
\subsubsection{Perturbation equations with source}
%
Scalar, vector and gravitational perturbations of the Kerr metric are  described by the Teukolsky equation\,\cite{teukolsky_perturbations_1973}:
%
\begin{widetext}
\begin{equation}
\begin{split}
\left[\frac{(r^2+a^2)^2}{\Delta}-a^2\sin^2\theta\right]\frac{\partial^2 \psi}{\partial t^2}+\frac{4Mar}{\Delta}\frac{\partial^2 \psi}{\partial \phi \partial t}+\left[\frac{a^2}{\Delta}-\frac{1}{\sin^2\theta}\right]\frac{\partial^2 \psi}{\partial \phi^2}-\Delta^{-s}\frac{\partial}{\partial r}\left(\Delta^{s+1}\frac{\partial \psi}{\partial r}\right)-\frac{1}{\sin\theta}\frac{\partial}{\partial \theta}\left(\sin\theta \frac{\partial \psi}{\partial \theta}\right)\\
-2s\left[\frac{a(r-M)}{\Delta}+\frac{i\cos\theta}{\sin^2\theta}\right]\frac{\partial \psi}{\partial \phi}-2s\left[\frac{M(r^2-a^2)}{\Delta}-r-ia\cos\theta\right]\frac{\partial \psi}{\partial t}+(s^2\cot^2\theta-s)\psi=4 \pi\Sigma T\,,
    \end{split}
\label{Teukolsky_Master_Equation}
\end{equation}
\end{widetext}
where the field $\psi$ identifies the type of perturbation, and $s=0,1,-2$ stands for scalar, vector and tensor modes, respectively. In our case we have
\begin{align}
    \psi(s=0)=\varphi \qquad \ \psi(s=-2)=(r-ia \cos\theta)^4 \bm{\psi}_4\,,
\end{align}
where $\bm{\psi}_4$ is a Weyl scalar. The source term is given by
\begin{equation}
    T(t,r,\theta,\phi)=-\frac{ d\  m_p}{\dot t \sin\theta} \delta(r-r(t))\delta(\theta-\theta(t))\delta(\phi-\phi(t))\ .
\label{J}
\end{equation}
where $\dot{t}$ is $dt/d\tau$ given in Eq. \eqref{tGeoEquation}. 
The Teukolsky equation is separable (see\,\cite{chandrasekhar_mathematical_2009} and references therein), in terms of an orthonormal set of angular functions, the spin-weighted spheroidal harmonics ${_s}S^\omega_{\ell m}(\theta)$\,\cite{teukolsky_perturbations_1973,Goldberg:1966uu}. By expanding the
field $\psi$ and the source term as
\begin{align}
   \psi(t,r,\theta,\phi)&= \sum_{\ell m}\int {_s}R^\omega_{\ell m}(r) {_s}S^\omega_{\ell m}(\theta) e^{im\phi-i\omega t}\dd \omega\ ,\label{psi_Harmonics_Decomposition}\\
  4\pi \Sigma T&=\sum_{\ell m}\int {_s}\mathcal{T}^\omega_{\ell m}(r){_sS}^\omega_{\ell m}(\theta)e^{im\phi-i\omega t}\dd \omega\,,\label{T_Harmonics_Decomposition}
\end{align}
it reduces to a decoupled set of ordinary differential equations.

Hereafter we focus on scalar perturbations only, i.e. we fix \footnote{For sake of simplicity we drop the subscript $s$ form the radial and the angular functions.} $s=0$, referring the reader to \cite{hughes_evolution_2008} for tensor modes.
The decoupled equations for the radial 
functions then read:
\begin{equation}
         \frac{\dd}{\dd r}\left(\Delta\frac{\dd}{\dd r} R_{\ell m}^\omega\right)+\Biggl(\frac{\kappa^2}{\Delta}-\lambda_{\ell m}\Biggr)R_{\ell m}^\omega=\source\,.\label{RTeukolsky_Equation}
\end{equation}
The spin-weighted spheroidal harmonics are the solutions of the equation:
\begin{align}
        \frac{1}{\sin\theta}\frac{\dd}{\dd\theta}\biggl(\sin\theta\frac{\dd\  S^\omega_{\ell m}}{\dd\theta}\biggr)+\biggl(a^2\omega^2\cos^2\theta-\frac{m^2}{\sin^2\theta}+\nonumber\\
        +\lambda_{\ell m}-a^2\omega^2-2am\omega\biggr)S^\omega_{\ell m}=0\,.\label{eq:spheroidal_harmonics}
\end{align}
Here $\lambda_{\ell m}$ is the eigenvalue of the spin-weighted spheroidal harmonic $S^\omega_{\ell m}$ and $\kappa(r)=(r^2+a^2)\omega-m a$.
We look for the solutions of Eq.~\eqref{RTeukolsky_Equation} with outgoing (ingoing) wave boundary conditions at infinity (horizon).
We hence follow the Green functions approach, by first solving the associated homogeneous problem, and then integrating the solutions over the source term. 

In order to solve the homogeneous equation, it is first convenient to redefine $R_{\ell m}^{\omega}$ as
\begin{equation}
    Y_{\ell m}^\omega(r)=(r^2+a^2)^{1/2} R_{\ell m}^{\omega}(r)\ ,\label{eq:YR}
\end{equation}
such that Eq.~\eqref{RTeukolsky_Equation} becomes 
\begin{equation}
    \frac{\dd^2Y^\omega_{\ell m}}{\dd r^{\star 2}}+\left[\frac{\kappa^2-\lambda_{\ell m}\Delta}{(r^2+a^2)^2}-G^2-G_{,r^\star}\right]Y_{\ell m}^\omega=T^\omega_{Y\ell m}\ ,\label{Yeq}
\end{equation}
where where $r^\star$ is the tortoise coordinate $\frac{dr^\star}{dr}=\frac{r^2+a^2}{\Delta}$
\cite{teukolsky_perturbations_1973}, $G= r\Delta/(r^2+a^2)^2$ and $T^\omega_{Y\ell m}$ are the coefficients of
\begin{equation}
    T_Y(t,r,\theta,\phi)=\frac{\Delta(r)}{(r^2+a^2)^{\frac{3}{2}}}T(t,r,\theta,\phi)\, 
\label{chi}
\end{equation}%
expanded as in\,\eqref{T_Harmonics_Decomposition}.

The homogeneous problem of Eq.~\eqref{Yeq} admits two independent solutions, with those being either purely outgoing at infinity ($Y_{\ell m\omega}^{+}$), or purely ingoing at the horizon ($Y_{\ell m\omega}^{-}$), given by:
\begin{equation}
       Y_{\ell m\omega}^{+}(r\rightarrow \infty)\sim e^{\mp i\omega r} \ \ ,\  
        Y_{\ell m \omega}^{-}(r\rightarrow r_H)\sim e^{\pm ip_\omega r^*}\ ,
\end{equation}
where $p_{\omega}=\omega-m a/2Mr_H$ and $r_H=M+\sqrt{M^2-a^2}$ is the radial coordinate at the event horizon. From Eq.~\eqref{eq:YR} we can also compute the asymptotic behavior of 
$R_{\ell m}^{\omega}$

\begin{align}
&R_{\ell m\omega}^-\sim\begin{cases}
    e^{-i p_\omega r^*}\quad &r\rightarrow r_H\\
    A_{in}r^{-1} e^{-i \omega r^*}+A_{out} r^{-1} e^{i \omega r^*} \quad & r\rightarrow \infty
\end{cases}\ ,\\
&R_{\ell m\omega}^+\sim\begin{cases}
    B_{in} e^{-i p_\omega r^*}+B_{out} e^{i p_\omega r^*} \quad &r\rightarrow r_H\\
    e^{i \omega r} \quad &r\rightarrow \infty
\end{cases}\ .
\end{align}
These functions are defined modulo an overall constant, which is irrelevant since it cancels in the final expression for non-homogeneous solutions\,\cite{hughes_evolution_2008}. The full solution of Eq.~\eqref{RTeukolsky_Equation} is then given by
\begin{equation}
 R_{\ell m}^{\omega}=Z^{-}_{\ell m\omega}(r)R^{-}_{\ell m\omega}(r)+Z^{+}_{\ell m\omega}(r)R^+_{\ell m\omega}(r)\ ,
\label{scalar_sol}
\end{equation}
where 
\begin{align}
    Z^{+}_{\ell m\omega}(r)&=\frac{1}{W}\int^{r}_{r_H}\frac{\rho^2+a^2}{\Delta(\rho)}\dd \rho\  Y^-_{\ell m\omega}(\rho)T^\omega_{Y\ell m}(\rho),\label{Z_infty}\\
     Z^{-}_{\ell m\omega}(r)&=\frac{1}{W}\int^{\infty}_{r}\frac{\rho^2+a^2}{\Delta(\rho)}\dd \rho\ Y^+_{\ell m\omega}(\rho)T^\omega_{Y\ell m}(\rho)\label{Z_H}\ ,
\end{align}
and $W= Y_{\ell m\omega}^-Y^+_{\ell m\omega,r}-Y^-_{\ell m\omega,r}Y^+_{\ell m\omega}$ is the Wronskian. The stress-energy tensor components $T^\omega_{Y\ell m}$ are given by
\begin{equation}
    T^\omega_{Y\ell m}(r)=
    -2 d m_p\int\dd t\frac{\Delta\delta[r-r(t)]}{(r^2+a^2)^{\frac{3}{2}}\dot t} e^{i[\omega t-m\phi(t)]}S^*_{\ell m}[\theta(t)]
\label{SourceY}
\end{equation}
where $S^*_{\ell m}$ is the complex conjugate of $S_{\ell m}$. For circular orbits $r(t)=r_0$, and we can use Eq.~\eqref{phi_thetaHarmonics_decomposition} to write $\exp[i m \phi(t)]$ as a series of harmonics in $\theta$ \cite{hughes_evolution_2008}. We define the function 
\begin{equation}
\begin{split}
      H_{\ell m}[r_0,\theta (t)]&= I_{\ell m}[r_0,\theta(t)]e^{im(\Omega_{\phi}t-\phi(t))}=\\
  =& \sum_{k=-\infty}^{\infty}H_{\ell mk}(r_0)e^{-i k \Omega_\theta t}\ ,
\label{H_definition}  
\end{split}
\end{equation}
where
\begin{equation}
    I_{\ell m}[r_0,\theta(t)]=-\frac{4\pi dm_p\Delta_0}{(r_0^2+a^2)^{\frac{3}{2}}}\frac{S^*_{\ell m}[\theta(t)]}{\dot t}\ ,
    \label{I_definition}
\end{equation}
and
\begin{equation}
H_{\ell mk}(r_0)=\frac{1}{T_{\theta}}\int_{0}^{T_\theta}\dd t \ H_{\ell m}(r_0,\theta(t))e^{ik\Omega_\theta t}\ .
\end{equation}
To avoid singularities in the domain of integration, we change variable $t\rightarrow \chi$. Then, $H_{\ell mk}$ reads
\begin{equation}
\begin{split}
     H_{\ell mk}(r_0)=&-\frac{4\pi dm_p\Delta_0}{T_\theta(r_0^2+a^2)^{\frac{3}{2}}}\\
     &\int_{0}^{2\pi}\dd\chi \frac{\gamma+a^2 Ez(\chi)}{\sqrt{\beta(z_+-z(\chi))}} \ \frac{S^*_{\ell m}[\theta(\chi)]}{\dot t} \\
     &\exp [i\omega_{mk} t(\chi)-im\phi(\chi)]\ ,
\end{split}
\end{equation}
with $\omega_{mk}= k\Omega_{\theta}+m\Omega_{\phi}$. Using the definitions introduced above the source term can be recast in the following form 
\begin{equation}
  T^\omega_{Y\ell m}(r)=\sum_{k=-\infty}^{\infty}\delta(r-r_0)\delta(\omega-\omega_{mk})H_{\ell mk}(r_0)\ .
\end{equation}
In the same way, it is convenient to decompose non-homogeneous solutions $Z^{\pm}_{\ell m\omega}$ as 
\begin{equation}
Z_{\ell m\omega}^\pm=\sum_{k\in \mathbb{Z}}Z^\pm_{\ell mk}\delta(\omega-\omega_{mk})\ .
\end{equation}
Then, defining 
\begin{equation}
\begin{split}
    \mathcal{I}_{\ell mk}(r_0)&=\int_0^{2\pi}\ d\chi \frac{\gamma+a^2 Ez(\chi)}{\sqrt{\beta(z_+-z(\chi))}} \ \frac{S^*_{\ell m}[\theta(\chi)]}{\dot t}\\
    &\exp[i\omega_{mk} t(\chi)-im\phi(\chi)]\,,
\label{I_lmk}   
\end{split}
 \end{equation}
and
\begin{equation}
   C^\pm_{\ell mk}= \frac{-4\pi m_p}{T_\theta W}\frac{Y^{\mp}_{\ell m\omega}(r_0)}{\sqrt{r_0^2+a^2}} \mathcal{I}_{\ell mk}(r_0)\ ,
\label{CDefinition}\end{equation}
we get 
\begin{equation}
    Z_{\ell mk}^{\pm}(r)=d\mathrm{\Theta}(x_{\pm})C^\pm_{\ell mk}\label{Z_inf_Circular}
\end{equation}
where $\omega=\omega_{mk}$, $\mathrm{\Theta}(x)$ is the Heaviside function, $x_+= r-r_0$ and $x_-= -x_+=r_0-r$.

%
\subsubsection{Scalar fluxes}
%
The energy and angular momentum fluxes, at leading order in the mass ratio, can be extracted from the asymptotic value of the scalar field stress-energy tensor, which is computed in terms of the scalar field solution derived above.
Following \cite{warburton_self-force_2010}, we introduce
\begin{equation}
 \dot E_\pm= \frac{\dd E_{\pm}}{\dd t}= \mp\int d\Omega \ \Delta T_{rt}\,,
\label{Energy_Derivative}
\end{equation}
where the upper (lower) sign is referred to the emission at the infinity (horizon). By replacing 
the scalar field solution\,\eqref{psi_Harmonics_Decomposition} , \eqref{scalar_sol}, $\varphi=\psi(s=0)$, in $T^{scal}_{\mu\nu}$ (see Sec.\ref{subsec:fields}), and exploiting  the asymptotic behaviour of the radial solution and  the properties of the spheroidal harmonics, we find 
\be
\dot E_\pm=\frac{d^2}{16\pi}\sum_{\ell=0}^\infty\sum_{m=-\ell}^\ell\sum_{k=-\infty}^\infty \omega_{mk}p^\pm_{mk}|C^\pm_{\ell mk}|^2\ ,
\label{Energy_dot}
\eeq
where $p^+_{mk}= \omega_{mk}$, 
$p^-_{mk}=  p_{\omega_{mk}}$, with $C^\pm_{\ell mk}$ being defined in Eq.~\eqref{CDefinition} \cite{barsanti_scalar_2019}. The energy and the angular momentum fluxes for each mode $(\ell,m,k)$ are related by
\begin{equation}
\dot L_{\ell mk}=\frac{m}{\omega_{mk}}\dot E_{\ell mk}\ ,
\end{equation}
therefore $\dot L$ 
\begin{equation}
    \dot L= \frac{\dd L}{\dd t}=\frac{d^2}{16\pi}\sum_{\ell=0}^\infty\sum_{m=-\ell}^\ell\sum_{k=-\infty}^\infty m p^\pm_{mk} |C^\pm_{\ell mk}|^2\ .
\label{Angular_Momentum_dot}
\end{equation}
As expected, both $\dE$ and $\dL$ scale with the square of the scalar charge.
%
\subsubsection{Adiabatic variation of orbital parameters}\label{sec:Adiabatica_varaition_of_orbital_parameters}
%
The scalar emission affects the EMRI dynamics, making the system coalescing faster due to the extra leakage of energy. The total energy and angular momentum fluxes are then given by
\begin{equation}
\dC=\dC_\textnormal{grav}+\dC_\textnormal{scal}\ ,
\label{Cdot}\end{equation}
where $\mathcal{C}=\{E,L\}$.  Since, as discussed in Sec.\,\ref{subsec:geodesics}, we consider orbits with vanishing eccentricity (which is possible since circular geodesics remain circular during the inspiral\,\cite{kennefick_radiation-reaction-induced_1996}), they can  be described in terms of two orbital parameters. 
We choose the fixed radial coordinate $r=r_0$  and the angular variable $x$ \cite{hughes_adiabatic_2021}, which determines the inclination of the orbit:
\begin{equation}
x=\cos\theta_{\tn{inc}} \, ,\label{eq:defx}
\end{equation}
where $\theta_{\tn{inc}}=\pi/2-\mbox{sgn}(L)\theta_{\tn{min}}$. The angle $\theta_{\tn{inc}}$ is measured with respect to the equatorial plane, while $\theta_{\tn{min}}$ is measured with respect to the BH spin axis: $\sin^2\theta_{\tn{min}}=x^2$. We also remark that $\theta_{\tn{inc}}$ is acute (obtuse) for prograde (retrograde) orbits.
Since the geodesics  are circular, 
\begin{equation}
   \frac{\dd \mathcal{R}(r)}{\dd t}=\frac{\dd\mathcal{R}^\prime(r)}{\dd t}=0\,,
\end{equation}
where $\prime$ indicates derivative with respect to $r$.
The condition $\dd\mathcal{R}^\prime / \dd t=0$ is sufficient to determine $\dot r$. We get
\begin{equation}
\dot r=\frac{(2 aL r-4 r^3 E-4a^2 E r)\dot E+2a E r \dot L+(r-M)\dot K}{6 M r-6 r^2(1-E^2)- a^2-K+2 a^2 E^2- 2E La}\ ,
    \label{rdot}
\end{equation}
where $K= Q+(L-aE)^2$. The variation of $K$ can be computed starting from $\mathcal{R}(r)=0$, which leads to
\begin{equation}
    K=\frac{1}{\Delta}\left[E(r^2+a^2)-La\right]^2-r^2\ .
\label{K1}
\end{equation}
Using the geodesic equation, it can be shown that the derivative of $K$ with respect to $r$ vanishes for circular orbits, and then 
\begin{equation}
    \dot K=\frac{2(r^2+a^2)E-aL}{\Delta}\left[(r^2+a^2)\dot E-a\dot  L\right].
\label{Kdot}
\end{equation}
At each orbit, the value of the extremal polar angle $\theta_{\tn{min}}$ is given by the equation $\Theta^2(\theta_{\tn{min}})=0$.  
Changing variable to $\theta_{\tn{min}}\to x$ we have that during the inspiral the following equation holds:
\begin{equation}
   0= \Theta^2(x)=K-\left(\frac{L}{x}\right)^2+2 a E L -a^2 E^2 x^2-a^2x^2\,,
\label{Eq_start}
\end{equation}
which can be written in the form
\begin{equation}
    \frac{L}{x}=\sqrt{K+2 a E L -a^2(1-x^2)-a^2 E^2 x^2}\ ,
\label{L_x_ratio}
\end{equation}
where there is no ambiguity in the sign since $x$ has always the same sign as $L$.
Differentiating with respect to $t$, and solving for $\dot x$ we get\,\footnote{We have numerically checked that Eq.\,\eqref{xdot} is equivalent to the expression derived with an independent approach in \cite{hughes_adiabatic_2021} (see 
Eqs.~B3 and B4 within).}:
\begin{widetext}
\begin{equation}
\dot x=-\frac{x\dot K + 2 a x (L-a E x^2)\dot E+ 2\left(a E x-\sqrt{K+2 a E L- a^2(1-x^2)-a^2 E^2 x^2}\right)\dot L}{2\left[K+2aEL-a^2+2x^2a^2(1-E^2)\right]}\ .
\label{xdot}
\end{equation}
\end{widetext}
%

%
\subsection{Gravitational waveform}\label{sec:Gravitational_Waveform}
%
With  formalism developed in Sec.~\ref{sec:perturbations}, it is possible to determine the fully relativistic  gravitational waveform of the EMRI, solution of Einstein's equation~\eqref{einstein}\,\cite{drasco_gravitational_2006}. 
However, this approach is computationally expensive, and its implementation in current pipelines for LISA data analysis is a numerical challenge even in GR \cite{Katz:2021yft}.

We adopt therefore a simpler model, which suffices the purpose of assessing the impact of the scalar field on the EMRI waveform: the so-called numerical kludge waveform \cite{babak_kludge_2008}, which is based on the quadrupole approximation. In this setup the GW strain is given by
\begin{equation}
h_{ij}=\frac{2}{d_{\rm L}}\frac{d^2 I}{dt^2}\quad \ ,\quad 
I_{ij}=m_p z^{i}(t)z^z(t)\ ,\label{Quadrupole_approximation}
\end{equation}
where $z^i(t)$ is the worldline of the secondary in Cartesian spatial coordinates and $d_\tn{L}$ the luminosity distance. 
By integrating Eqs.~\eqref{tGeoEquation}-\eqref{VarphiGeoEquation},
~\eqref{t_Derivative_chi} and~\eqref{rdot}, 
with the energy and   
angular momentum fluxes computed in Sec.\ref{sec:perturbations} (Eq.\,\eqref{Cdot}), we find the evolution of the orbital elements of the secondary in Boyer-Lindquist coordinates $r(t), \chi(t)$ and $\phi(t)$. In the transverse-traceless gauge, the physical propagating degrees of freedom along radial direction from a source at an azimuthal angle $\vartheta$ and a polar angle $\Phi$ are given by $h^{\mbox{+}}= \frac{1}{2}(h^{\vartheta\vartheta}-h^{\Phi\Phi})$ and $\hcross= h^{\vartheta\Phi}$, where $h^{\mbox{+}}$ and $\hcross$ are the “plus" and “cross" waveform polarizations and
\begin{align}
h^{\vartheta\vartheta}=&\cos^2\vartheta\big[h^{xx}\cos^2\Phi
+h^{xy}\sin2\Phi+h^{yy}\sin^2\Phi\big]\nonumber\\
&+h^{zz}\sin^2\vartheta\sin2\vartheta[h^{xz}\cos\Phi+h^{yz}\sin\Phi]\ ,\label{eq:hthth}\\
h^{\Phi\vartheta}=&\frac{\cos\vartheta}{2}\big[2h^{xy}\cos2\Phi-h^{xx}\sin2\Phi+h^{yy}\sin2\Phi\big]\nonumber\\
&+\sin\vartheta\left[h^{xz}\sin\Phi-h^{yz}\cos\Phi\right],\\
    h^{\Phi\Phi}=&h^{xx}\sin^2\Phi-h^{xy}\sin2\Phi+h^{yy}\cos^2\Phi\ .\label{eq:hphph}
\end{align}
Hereafter, we assume binaries with $\vartheta=\pi/3$ and 
$\Phi=0$. 
\subsection{Numerical implementation}\label{subsec:num}
%
Numerical calculations of gravitational and scalar fluxes are performed using dedicated \texttt{Mathematica} packages. We compute geodesic quantities and homogeneous solutions to Teukolsky equations using the Black Hole Perturbation Toolkit (BHPT) \cite{noauthor_black_nodate}, while we have developed an independent code for integrations over the source terms.

In the next section we show results for a prototype EMRI with a primary BH having a spin parameter $a=0.95M$. Energy and angular momentum fluxes are computed on a rectangular grid $(y,x)$ populated by $41\times 11$  points evenly distributed, where $y\in[0,1]$ is defined as 
\begin{align}
    y(r,x)=& \frac{u(r,x)-u(r_{\tn{max}},x)}{u(r_{\tn{min}},x)-u(r_{\tn{max}},x)}\ ,\\
    u(r,x)=&\frac{1}{\sqrt{r-0.9 \,r_\textnormal{ISSO}(x)}}\ ,
\end{align}
and $r_\textnormal{ISSO}$ is the Innermost Spherical Stable Orbit (ISSO), which depends on $x$. We choose $r_{\tn{min}}=r_\textnormal{ISSO}+\delta r$ and $r_{\tn{max}}=r_\textnormal{min}+10 M$, where we  added the factor $\delta r=0.2$ to avoid singularities, and we remind that the variable $x\in[-1,1]$\,\eqref{eq:defx} is positive (negative) for  prograde (retrograde) orbits. 

Both gravitational and scalar fluxes are given as sums over the multipolar indices $\ell,m$ and $k$, as shown  in Eq.~\eqref{Energy_dot}. 
Summation on $k$ can be simplified by exploiting the symmetry properties of the solution
\begin{equation}
Z_{\ell-m-k}=(-1)^{\ell+k}Z_{\ell mk}^*\,,
\label{Zlmk}
\end{equation}
and using the fact that modes with $\ell+m+k=2n+1,\ n\in \mathbb{N}$ are vanishing (see Appendix \ref{sec:Appendix_kmodes}).
We truncate the (infinite) sum on $\ell$ and $k$ by adopting the accuracy criteria introduced in \cite{hughes_evolution_2008}: (i) we stop the series in $k$ when the energy fluxes satisfy the condition
\begin{equation}
    \dot E_{\ell mk}\le \epsilon_k\times\dot E_{\ell m}^\textnormal{leading term},
    \quad \epsilon_k\ll 1
\label{HughesCriterion_k}
\end{equation}
for $n_k$ times in a row; (ii) we truncate the series in $\ell$ when
\begin{equation}
    \dot E_{\ell}=\sum_{mk}\dot E_{\ell mk}\le \epsilon_\ell\times\dot E_{\ell}^\tn{leading term},\quad \epsilon_k \ll \epsilon_\ell \ll 1
\label{HughesCriterion_l}
\end{equation}%
for $n_\ell$ times in a row. We fix $\epsilon_\ell =10^{-3}$ and $\epsilon_k=\epsilon_\ell /10$ such that the relative error on the energy and angular momentum fluxes is $\lesssim 10^{-3}$. Moreover, we choose\footnote{These upper bounds are never reached with the choice $\epsilon_{\ell}=10^{-3}$.} $\l_{max}=k_{max}=20$.

With the values of the fluxes at each point of the grid, we can compute the change in the orbital elements through Eqs.~\eqref{rdot} and \eqref{xdot}. The right-hand side of such equations can be cast in order to isolate the GR and the scalar field contributions, which are numerically interpolated through \texttt{Mathematica}. 
The final set of coupled equations
\begin{align}
    \frac{\dd r(t)}{\dd t}&=\dot r_{\rm grav}[r(t),x(t)]+d^2\dot r_{\rm scal}[r(t),x(t)]\ ,
    \label{r_to_integrate}\\
    \frac{\dd x(t)}{\dd t}&=\dot x_{\rm grav}[r(t),x(t)]+d^2\dot x_{\rm scal}[r(t),x(t)]\ ,
    \label{x_to_integrate}
\end{align}
is then integrated with a suitable choice of 
the initial conditions. The solutions allows to compute $\chi(t)$ and $\phi(t)$, and the GW polarizations \eqref{eq:hthth}-\eqref{eq:hphph} in the time domain, where
\begin{align}
    h_{xx}&=r^2\left(\sin^2\chi+x^2\cos^2\chi\right)\cos^2\phi\ ,\nonumber\\
    h_{yy}&=r^2\left(\sin^2\chi+x^2\cos^2\chi\right)\sin^2\phi\ ,\nonumber\\
    h_{zz}&=r^2\left(1-x^2\right)\cos^2\chi\ ,\nonumber\\
    h_{xy}&=\frac{1}{2}\left(\sin^2\chi+x^2\cos^2\chi\right)r^2\sin2\phi\ ,\nonumber\\
    h_{xz}&=r^2\sqrt{1-x^2}\cos\chi\sqrt{\sin^2\chi+x^2\cos^2\chi}\cos\phi\ ,\nonumber\\
    h_{yz}&=r^2\sqrt{1-x^2}\cos\chi\sqrt{\sin^2\chi+x^2\cos^2\chi}\sin\phi\ .\nonumber
\end{align}
As a final step we perform a Discrete Fourier Transform of $h_{+,\times}$ to map the signal in the frequency space (see Paper II). 

We have tested our code reproducing results available in literature in GR \cite{hughes_evolution_2008}, finding an agreement on the GW fluxes up to machine precision. Moreover, to test the interpolation for $r(t)$ and $x(t)$ we have considered a smaller grid in the $(y,x)$ plane with 126 total points. We find an average relative difference with respect to values interpolated from the larger grid of $\sim 10^{-4}$. Such value increases up to $\sim 10^{-2}$ for orbital radii close to the plunge.

%
\section{Results}\label{sec:results}
%

\begin{figure}
    \centering
    \includegraphics[width=0.8\columnwidth]{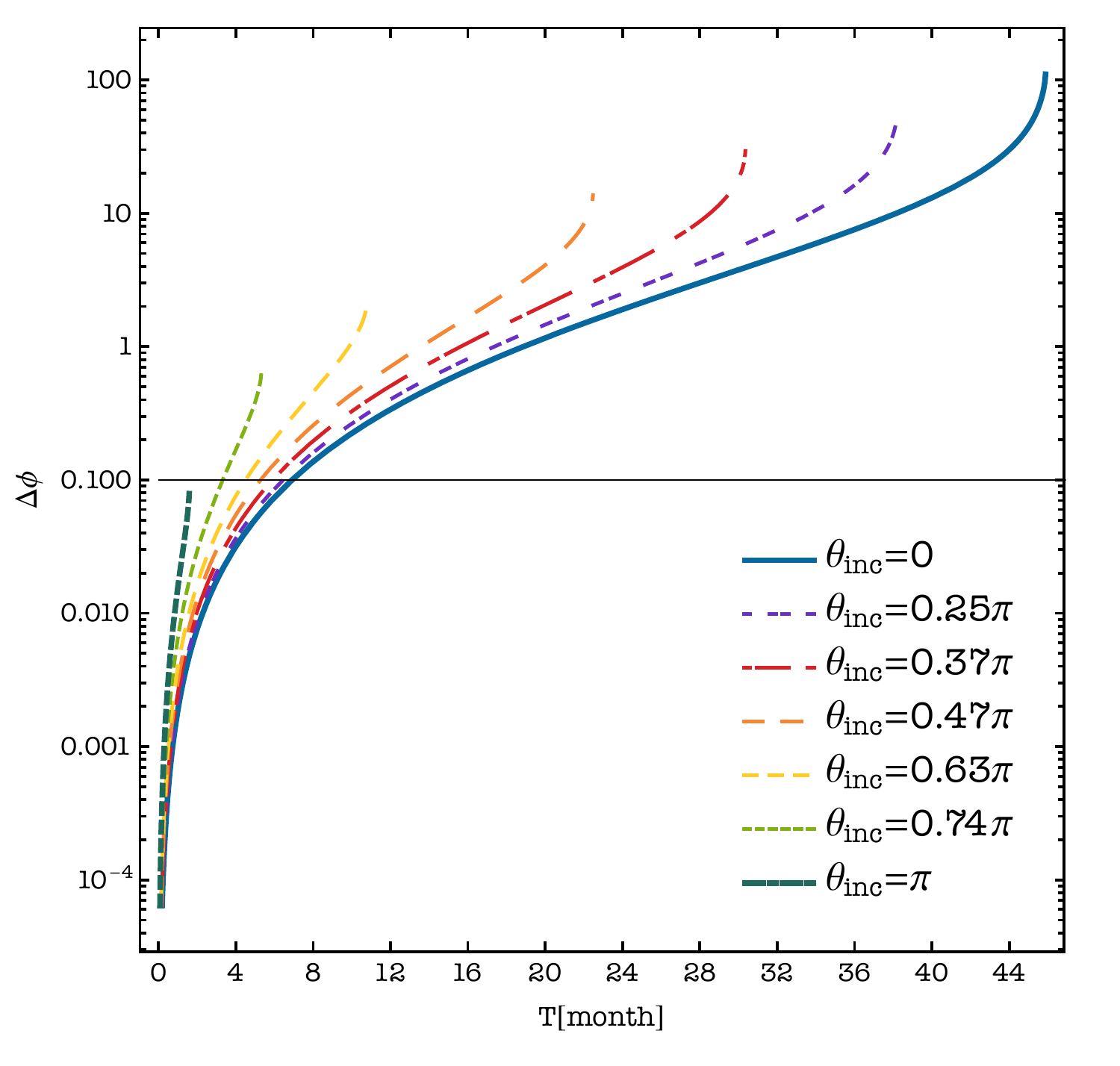}
      \caption{Quadrupolar dephasing as function of the observing time for different values of the initial inclination of the orbit. We fix the scalar charge to $d=0.01$. The horizontal line corresponds to the threshold value for detectability, $\Delta \phi=0.1$, for an EMRI observed by LISA with SNR$=30$ \cite{PhysRevLett.123.101103}.}\label{fig:dephasing}
\end{figure}

For a preliminary assessment of the effect of the orbital inclination on the detectability of the scalar charge, we  consider the quadrupolar dephasing $\Delta\phi= 2 \times \left[ \phi_{{\rm d}}(t)-\phi_{0}(t) \right]$, where $\phi_0=\phi_{\rm d=0}$, and the the two phases are computed with the same initial conditions.
In Fig.~\ref{fig:dephasing} we show $\Delta\phi$ as a function of the observing time, for $d=0.01$. Binaries evolve from an orbital separation $r_0=10M$ until the plunge, with different values of the initial inclination angle. We also show (horizontal line) the threshold value $\Delta\phi=0.1$, above which two signals observed by LISA with a signal-to-noise ratio (SNR) of $30$ are expected to be distinguishable \cite{PhysRevLett.123.101103}.
We can see that  after one year of observation, the dephasing increases above the distinguishability threshold for all values of $\theta_{\rm inc}$ except for $\theta_{\rm inc}=\pi$. This analysis confirms the results obtained in Papers I,II and \,\cite{barsanti_extreme_2022} for equatorial circular and eccentric orbits.

Fig.~\ref{fig:dephasing} also shows that, for a given time, $\Delta\phi$ increases for larger inclination angles, and it is maximum for retrograde configurations, i.e. $\theta_{\rm inc}\in[\pi/2,\pi]$.
Note that the \textit{total} dephasing, evaluated at the time of plunge, is instead larger for less inclined orbits. This is due to the specific setup of our analysis, which assumes the same initial separation for all systems. Indeed, EMRIs with $\theta_{\rm inc}\neq 0$ reach the plunge faster as the separatrix shrinks, resulting in an overall smaller number of accumulated cycles.

To obtain a more quantitative assessment of the detectability of scalar charge for inclined orbits, we investigate the faithfulness $\mathcal F$ between two waveforms in the frequency domain $\tilde{h}_1(f)$ and $\tilde{h}_2(f)$
\begin{equation}
\mathcal{F} [h_1,h_2]=\max_{t,\phi} \frac{\braket{h_1}{h_2}}{\sqrt{\braket{h_1}{h_1}\braket{h_2}{h_2}}},
\label{eq:faithfulness}
\end{equation}
where Eq.~\eqref{eq:faithfulness} is maximised over time 
and phase shifts \cite{Lindblom:2008cm}. The inner product $\braket{a}{b}$ is given by
\begin{equation}
    \braket{a}{b}=4\Re\int_{f_{min}}^{f_{max}} \frac{\tilde{a}(f)\tilde{b}^*(f)}{S_n(f)}\dd f \ ,
\end{equation}
and the $S_n(f)$ is the LISA noise Power Spectrum Density, also including the confusion noise produced by galactic white dwarf binaries \cite{Robson:2018ifk}. 
We set $f_{min}=10^{-4}$Hz, while $f_{max}$ corresponds to the orbital frequency at $r_{\textnormal{min}}$. As a rule of thumb, two signals with SNR$=30$ are distinguishable if ${\cal F}\lesssim 0.994$ \cite{Chatziioannou:2017tdw}\,.

To assess the convergence of our results, we have computed $\mathcal{F}$ for given $\theta_{\rm inc}$ and $d$ increasing the working precision. On average we find maximum deviations of the order of $5\%$, which we consider as systematic error of our calculations.

\begin{figure}
    \centering
    \includegraphics[width= 0.8\columnwidth]{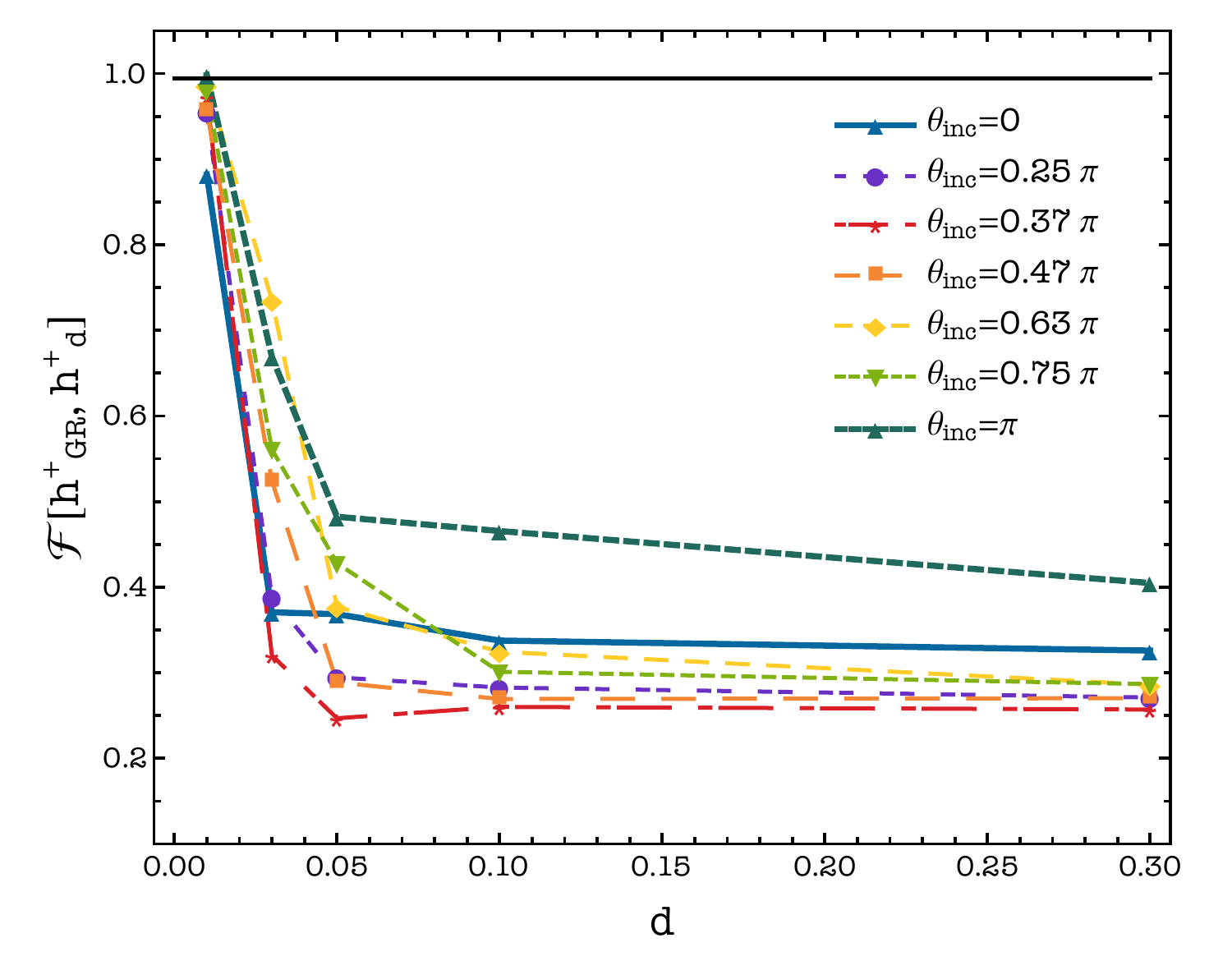}
    \caption{Faithfulness between the plus polarization of the GW signal computed in GR and in presence of a non-vanishing scalar charge $d$, as function of the scalar charge, for different values of the initial inclination angle. We assume one year of observation before the plunge. The horizontal line identifies the threshold below which the two signals are distinguished by LISA for an EMRI observed with SNR=30 \cite{Chatziioannou:2017tdw}.}\label{fig:faith_vs_charge}
\end{figure}

Fig.\,\ref{fig:faith_vs_charge} shows the faithfulness between the $h_+$ polarization computed in GR and in presence of a non-vanishing scalar charge $d$, for one year of observation time until the plunge, 
and different initial inclination angles $\theta_{\tn{inc}}$. 
Note that in this case the initial orbital separation of the binary is not the same in the different models: it is a function of $d$ and $\theta_{\rm inc}$. 
The results shown in Fig.\,\ref{fig:faith_vs_charge} confirm the dephasing analysis. They also show that for $d\gtrsim0.05$ the faithfulness sharply drops from one, for all configurations considered, and saturates around ${\cal F}\sim0.4$. While inclined configurations yield smaller values of ${\cal F}$, this trend changes for retrograde orbits with $\theta_{\rm inc}> 0.74\pi$, since these system plunge faster allowing for a shorter frequency integration. 

\begin{figure}[t]
    \centering
    \includegraphics[width= 0.9\columnwidth]{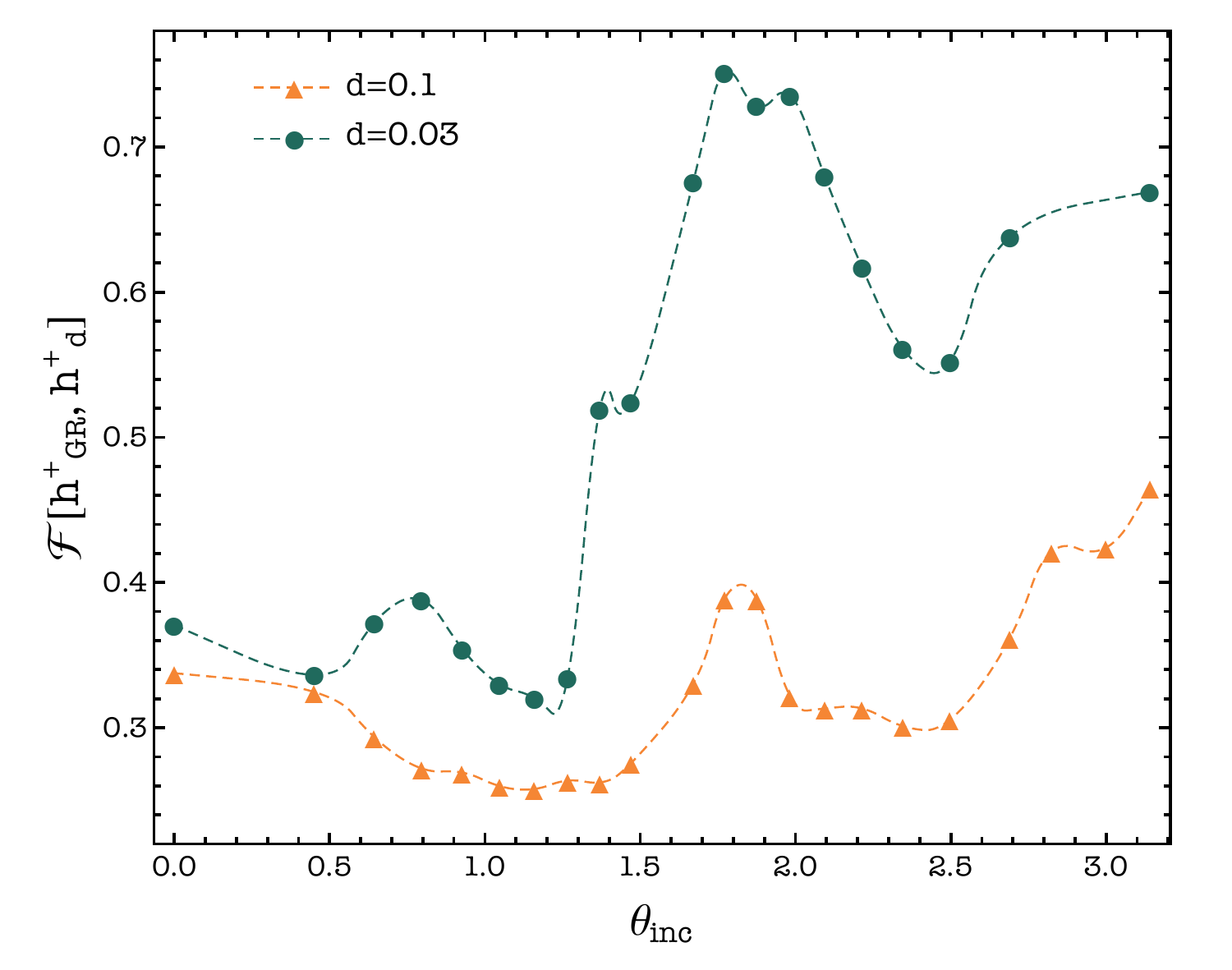}
    \caption{ 
    Same as Fig.~\ref{fig:faith_vs_charge} but as function of the initial inclination, for different values of the scalar charge.} \label{fig:faith_vs_incl}
\end{figure}

Finally, in Fig.~\ref{fig:faith_vs_incl} we show the faithfulness as a function of $\theta_{\rm inc}$,  for $d=0.03$ and $d=0.1$. 
This analysis shows that the faithfulness is significantly smaller for prograde orbits than for retrograde ones. For orbits of the same kind (either prograde or retrograde), the faithfulness has a mild dependence on the inclination angle.
This behaviour holds for different values of the charge.

%
\section{Conclusions}\label{sec:conclusions}
%
EMRIs are among the primary targets of the future space interferometer LISA. Their long inspiral evolution allows to estimate the source parameters with exquisite precision, rendering such binaries golden tools for precise tests 
of gravity in the strong field regime.

Modelling EMRIs beyond GR is still in its early stages, although a new framework to describe such binaries within Self-Force and in theories of gravity with extra scalar fields, has recently been developed \cite{Spiers:2023cva}.  
This approach is theory-agnostic at the adiabatic order in 
the mass ratio, with changes in the binary evolution 
uniquely determined by the scalar charge of the EMRI 
secondary, $d$. The approach builds upon a series of 
recent works, which determined the adiabatic evolution of 
EMRI on equatorial circular and eccentric orbits for 
massless fields \cite{Maselli:2020zgv,barsanti_scalar_2019}, and for equatorial circular inspirals and massive scalars \cite{barsanti_detecting_2022}.
In this paper have extended the description of EMRIs 
with massless scalar fields to inclined circular orbits around Kerr black holes, for both prograde and retrograde trajectories. We have computed the gravitational and scalar fluxes, which drive the EMRI adiabatic evolution, and assessed the relevance of the orbital inclination on the detectability of $d$.

We have computed the dephasing induced by the presence of the scalar charge for different inspirals up to the plunge. 
This analysis suggests that LISA could be able 
to detect charges as small as $d=0.01$.

We have then performed a more rigorous analysis, based on the faithfulness computed between gravitational waveforms in GR  and with a non-vanishing scalar charge. 
We find that scalar charges with $d \gtrsim 0.05$ could lead to distinguishable signals after one year of observation in the LISA band, consistently with the results of Papers I-III. We also find that the faithfulness is significantly smaller for prograde orbits than for retrograde ones. By focusing on orbits which are either prograde or retrograde, we find that the faithfulness mildly decreases (increases) for larger values of the initial inclination angle for prograde (retrograde) orbits. This suggests that inclined prograde orbits could leave a larger imprint of the charge on the emitted waveform.

Both the dephasing and the faithfulness provide only preliminary indications  on the actual impact of GR corrections on the EMRI waveform, as they do not take into account correlations among the source parameters.
A fully Bayesian analysis based on Monte Carlo Markov Chain simulations with Fast EMRI Waveforms \cite{Katz:2021yft, Chua:2020stf} is currently in preparation \cite{sperinew}. We are also planning to extend our formalism to generic configurations, i.e. treating EMRIs on eccentric, inclined orbits. This will complete the description of EMRIs 
with scalar fields at the adiabatic order. 
Efforts to include post-adiabatic corrections \cite{Spiers:2023cip}, including spin and 
dipole contributions from the secondary 
\cite{Lestingi:2023ovn} are underway and 
will require a longer path.

\newpage
\appendix
%
\section{Condition for non vanishing k modes}\label{sec:Appendix_kmodes}
In this Appendix we show that, as anticipated in Sec.~\ref{subsec:num}, modes of the solution of the Teukolsky equation with odd $\ell+m+k=2n+1$ identically vanish. To this aim, we shall show that the integral  in Eq.\,\eqref{I_lmk},
\begin{equation}
\begin{split}
    \mathcal{I}_{\ell mk}(r_0)&=\int_0^{2\pi}\ d\chi \frac{\gamma+a^2 Ez(\chi)}{\sqrt{\beta(z_+-z(\chi))}} \ \frac{S^*_{\ell m}[\theta(\chi)]}{\dot t}\\
    &\times e^{i(k\Omega_\theta+m\Omega_{\phi}) t(\chi)-im\phi(\chi)}\,,
\label{I_lmk2}   
\end{split}
 \end{equation}
vanishes when $\ell+m+k$ is odd; this leads to the vanishing of the corresponding function $Z_{\ell mk}$\,\eqref{Z_inf_Circular}.

Let us consider the symmetry properties of the integrand of\,\eqref{I_lmk2} under reflection with respect to the equatorial plane $\theta=\chi=\pi/2$, i.e. for the transformation $\theta\to\pi-\theta$, which corresponds to $\chi\to\pi-\chi$ (see Eq.\,\eqref{chi_Definition}). Since $z=\cos^2\theta$, $\frac{\gamma+a^2 Ez(\chi)}{\sqrt{\beta(z_+-z(\chi))}}$ is invariant for this transformation.
The properties of the spheroidal harmonics\,\cite{Goldberg:1966uu} 
for even spin $s$ (we are considering the cases $s=0,2$) imply that
\begin{equation}
S^*_{\ell m}[\pi-\theta]=(-1)^{\ell+m}S^*_{\ell m}[\theta].
\end{equation}

Finally, since the particle employ half-period $T_\theta$ to reach the opposite position with respect to the equatorial plane, we find:
\begin{align}
t(\pi-\chi)-t(\chi)&=\frac{1}{2}{T}_\theta=\frac{\pi}{\Omega_\theta}\nonumber\\
\phi(\pi-\chi)-\phi(\chi)&=\frac{1}{2}\bar\phi=\pi\frac{\Omega_\phi}{\Omega_\theta}
\end{align}
thus
\begin{align}
&(k\Omega_{\theta}+m\Omega_\phi)t(\pi-\chi)-m\phi(\pi-\chi)\nonumber\\
&-(k\Omega_{\theta}+m\Omega_\phi)t(\chi)+m\phi(\chi)\nonumber\\
=&(k\Omega_{\theta}+m\Omega_\phi)
\frac{\pi}{\Omega_\theta}-m\pi
\frac{\Omega_\phi}{\Omega_\theta}
=k\pi\,.
\end{align}

Therefore, for $\theta\to\pi-\theta$
\begin{equation}
e^{i(k\Omega_\theta+m\Omega_{\phi}) t-im\phi}\to(-1)^k
e^{i(k\Omega_\theta+m\Omega_{\phi}) t-im\phi}\,.
\end{equation}

Putting all together, we get that for $\theta\to\pi-\theta$,
the integrand in Eq.\,\eqref{I_lmk2} is multiplied by a factor $(-1)^{l+m+k}$.

Therefore, if $l+m+k$ is odd, the integrand is antisymmetric with respect to equatorial reflection 
in its integration domain, and the integral \eqref{I_lmk2} is then vanishing.

\section{Non-equatorial geodesics in terms of elliptic functions}
\label{app:elliptic}
Eqs.\,\eqref{t_Derivative_chi}-\eqref{Phi_Derivative_chi} 
can be integrated using elliptic functions. Their general solution is:
\begin{align}
\begin{split}
         t(\chi) =\frac{\gamma}{\sqrt{\beta z_+}}\left[\tilde{{\cal K}}-{\cal F}\left(\frac{\pi}{2}-\chi,\tilde{z}\right)\right]+\\
   +a^2E\sqrt{\frac{z_+}{\beta}}\bigg[{\cal E}\left(\frac{\pi}{2}-\chi,\tilde{z}\right)-\tilde{\cal E}+\\
   +\tilde{{\cal K}}-{\cal F}\left(\frac{\pi}{2}-\chi,\frac{z_-}{z_+}\right)\bigg],
\end{split}
\label{t_chi}
\end{align}
\begin{align}
   \begin{split}
       \phi(\chi)=\frac{L}{\sqrt{\beta z_+}}\left[\tilde \Pi-\Pi\left(\frac{\pi}{2}-\chi,z_-,\frac{z_-}{z_+}\right)\right]+\\
    +\frac{\delta}{\sqrt{\beta z_+}}\left[{\tilde{\cal K}}-{\cal F}\left(\frac{\pi}{2}-\chi,\frac{z_-}{z_+}\right)\right].
   \end{split}
\label{phi_chi}
\end{align}
where ${\cal K}$ is the complete elliptic integral of the first kind, ${\cal F}$, ${\cal E}$, ${\Pi}$ are the incomplete elliptic integrals of first, second and third kind, respectively,
$\tilde{z}=z_-/z_+$, $\tilde{{\cal K}}= {\cal K}(\tilde{z})$, 
$\tilde{\cal E}={\cal E}\left(\pi/2,\tilde{z}\right)$, $\tilde{\Pi}=\Pi\left(\pi/2,z_-,\tilde z\right)$\,\cite{abramowitz+stegun}.
Although elliptic functions are usually defined in the domain $[0,\pi/2]$, they can be straightforwardly extended to $0\le \chi \le 2\pi$.

\bibliography{Bibliografia}

\end{document}